\documentclass[preprint,
amssymb, amsmath,
 aip,jcp]{revtex4-1}

\usepackage{bm}%
\usepackage[bookmarks=false]{hyperref}
\usepackage{graphicx}
\usepackage{braket}
\usepackage{color,soul}%
\usepackage{natbib}%
\usepackage{float}
\usepackage{diagbox}

\expandafter\ifx\csname package@font\endcsname\relax\else
 \expandafter\expandafter
 \expandafter\usepackage
 \expandafter\expandafter
 \expandafter{\csname package@font\endcsname}%
\fi

\begin{document}

\title{Energy transfer under natural incoherent light: Effects of asymmetry on efficiency}%

\author{Kenneth A. Jung}%
\email{kenneth.jung@utoronto.ca}
\affiliation{Chemical Physics Theory Group, Department of Chemistry, and Center for Quantum Information and Quantum Control, University of Toronto,Toronto,Ontario M5S 3H6, Canada}

\author{Paul Brumer}
\email{paul.brumer@utoronto.ca}
\affiliation{Chemical Physics Theory Group, Department of Chemistry, and Center for Quantum Information and Quantum Control, University of Toronto,Toronto,Ontario M5S 3H6, Canada}

%\date{\today}

\begin{abstract}
The non-equilibrium stationary coherences that form in donor-acceptor systems are investigated to determine their relationship to the efficiency of energy transfer to a neighboring reaction center. It is found that the effects of asymmetry in the dimer are generally detrimental to the transfer of energy. Four types of systems are examined, arising from combinations of localized trapping, delocalized (Forster) trapping, eigenstate dephasing and site basis dephasing. In the cases of site basis dephasing the interplay between the energy gap of the excited dimer states and the environment is shown to give rise to a turnover effect in the efficiency under weak dimer coupling conditions. Furthermore, the nature of the coherences and associated flux are interpreted in terms of pathway interference effects. In addition, regardless of the cases considered, the ratio of the real part and the imaginary part of the coherences in the energy-eigenbasis tends to a constant value in the steady state limit. 
\end{abstract}

\maketitle

\section{Introduction}
\label{sec:intro}
The existence of light induced coherences in biological systems and their effects on efficiency has been a topic of considerable interest.\cite{Fleming2011,Ball2011,Wu2012} Most of these studies have focused on excitation with ultra-fast coherent light sources, \cite{Engel2007,Ishizaki2012} which is not typical of natural conditions. Rather, incident sunlight is incoherent and provides a constant intensity over the excitation period.\cite{Jiang1991,Brumer2018} Studies of energy transfer in donor-acceptor(DA) systems under incoherent radiation\cite{Manal2010,Tscherbul2018,Manzano2013,Dodin2019,Dodin2016} have provided evidence that it is possible for long time coherences to occur under the influence of a natural incoherent pumping source when the system is coupled to a secondary bath. 

There have been a variety of approaches to describe the effects of natural radiation on quantum subsystems. Many of the early studies focused on the incoherent light-matter interactions using Bloch-Redfield theory, although a white noise description\cite{olina2014natural} has also been developed. These results were then extended to include effects of trapping and bath dephasing through the Lindblad formalism, which treats the incoherent light in the secular approximation.\cite{LenMontiel2014} However, describing light-matter interactions at the secular level treats the populations and coherences on different footings, and important effects such as Fano coherences cannot be observed.\cite{Pachn2017} A recent study\cite{Tscherbul2018} remedied this situation by adopting a hybrid approach that combines the non-secular description of the Bloch-Redfield treatment with a Lindblad description for the rest of the environment. With the use of this formalism analytic results were obtained for a symmetric DA system in the steady state limit. As insightful as this study is it focused on the idealized, symmetric case which is not typical of what is found in nature. It is to this end that we study the interplay of weak incoherent light and a phonon bath in an asymmetric DA model, building upon the previous master equation approaches, to create a general description that incorporates the important features of the light, the effects of a thermal bath, energy relaxation, and the role of an energy bias in the system. The most obvious goal for this study is physical insight, since one would not expect most naturally occurring DA systems to be symmetric. 

The layout of the paper is as follows. We first discuss, in Section \ref{sec:Theory}, a generalization to the master equations that incorporates an energy gap in the donor-acceptor system Hamiltonian as well as the metrics used to quantify the energy transfer. Two issues are emphasized: the relation of the trapping efficiency to the flux in the steady state limit, and the nature of the flux as an interference phenomenon. Numerical results obtained from these master equations by evolving them to the naturally relevant steady state limit are presented and discussed in Section \ref{sec:Results}. Section \ref{sec:Con} summarizes our findings.

%Figure 0
%--------------------------------------------------------------------------------------------
\begin{figure}[t]
\centering
\includegraphics[width=0.8\textwidth]{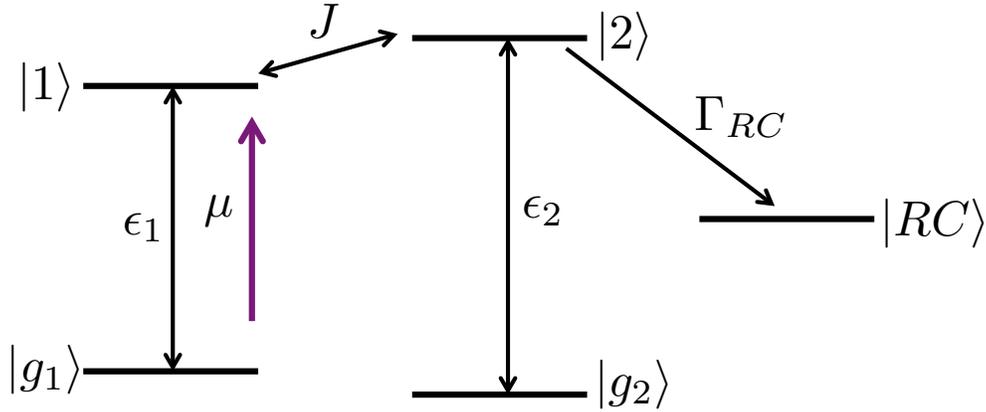} %figure name goes here
%\vspace*{-16mm}
\caption{The donor-acceptor exciton Hamiltonian. The dipole (in purple) is aligned with the donor system. The excited state of the acceptor transfers energy to the reaction center site at a rate $\Gamma_{RC}$.}
\protect\label{fig:0}
\end{figure}
%--------------------------------------------------------------------------------------------

\section{Theory}
\label{sec:Theory}

\subsection{Model system for photosynthetic energy transfer}
\label{sec:Model}
The model system used in this study is that presented in Ref.\ \citenum{Tscherbul2018}, i.e., a simple donor-acceptor (DA) exciton system
\begin{equation}
H_s = \sum^2_{k=1} \epsilon_k | k\rangle \langle k | + J\left( | 1\rangle \langle 2 | + | 2\rangle \langle 1 |  \right),
\end{equation}
with on-site energies $\epsilon_k$ and site hopping coefficient $J$. However, unlike Ref.\ \citenum{Tscherbul2018} the system here is asymmetric, i.e. $\epsilon _1\neq\epsilon _2$.  The dipole $\hat{\mu} = \mu_1|g_1\rangle \langle 1|$ is assumed to be aligned on the donor site, see figure \ref{fig:0}. The DA system interacts with an incoherent radiation field and is coupled to a reaction center. The description of the reduced density matrix is that of a hybrid Bloch-Redfield-Lindblad approach, meaning that the light-matter interactions will be treated using standard non-secular Markovian\cite{Jesenko2013} optical master equations\cite{Tscherbul2014,Tscherbul2015,Dodin2016,Kozlov2006} and the effects of the bath, recombination of excitons, and the transfer of energy to the reaction center is treated as a dephasing of the system through the use of Lindblad operators.\cite{Lindblad1976,LenMontiel2014} Under these assumptions the density matrix evolves according to
\begin{equation}
\label{ed:MEE}
\dot{\rho} = (L_0 + L_{\textrm{deph}} + L_{\textrm{rec}} + L_{\textrm{RC}} + L_{\textrm{rad}} )\rho,
\end{equation}
where $L_0\rho=-i[H_s,\rho]$.
The effects of the DA system coupled to the surrounding environment are given by
\begin{equation}
\label{eq:Ldephase}
L_{\textrm{deph}}\rho = 2\gamma_d\sum^2_{k=1} |k\rangle\langle k| \rho |k\rangle\langle k| -\frac{1}{2}[|k\rangle\langle k|, \rho]_+,
\end{equation}
where $[A,B]_+$ is the anticommutator of $A$ and $B$ and $\gamma_d$ is the dephasing rate. The term 
\begin{equation}
\label{eq:Lrecom}
L_{\textrm{rec}}\rho = 2\Gamma\sum^2_{k=1} |0\rangle\langle k| \rho |k\rangle\langle 0| -\frac{1}{2}[|k\rangle\langle k|, \rho]_+,
\end{equation}
accounts for the recombination of the excitons, and transfer from the acceptor site to the reaction center $|RC\rangle$ is described by
\begin{equation}
\label{eq:LRC}
L_{\textrm{RC}}\rho = 2\Gamma_{RC} \left(|RC\rangle\langle 2| \rho |2\rangle\langle RC| -\frac{1}{2}[|2\rangle\langle 2|, \rho]_+\right).
\end{equation}
The Bloch-Redfield equations for the light-matter interaction are most easily described in the energy-eigenbasis\cite{Breuer2007,Tscherbul2018} and are given by the following set of elements of $L_{\textrm{rad}}\rho$:
\begin{eqnarray}
\label{eq:Lrad}
(L_{\textrm{rad}}\rho)_{e_{\pm}e_{\pm}}  &=& -(r_{e_{\pm}} + \gamma_{e_{\pm}})\rho_{e_{\pm}e_{\pm}} + r_{e_{\pm}}\rho_{gg} - (\sqrt{r_{e_{+}}r_{e_{-}} } + \sqrt{\gamma_{e_{+}}\gamma_{e_{-}} })\rho^{R}_{e_{+}e_{-}} \nonumber \\
(L_{\textrm{rad}}\rho)_{e_{+}e_{-}} &=& -\frac{1}{2}(r_{e_{+}} + \gamma_{e_{+}} +r_{e_{-}} + \gamma_{e_{-}} + 2i\Delta)\rho_{e_{+}e_{-}} + \sqrt{r_{e_{+}}r_{e_{-}} }\rho_{gg} \\
& & - \frac{1}{2}(\sqrt{r_{e_{+}}r_{e_{-}} } + \sqrt{\gamma_{e_{+}}\gamma_{e_{-}} })(\rho_{e_{+}e_{+}}+\rho_{e_{-}e_{-}}), \nonumber
\end{eqnarray}
where $\Delta = \sqrt{(\epsilon_1-\epsilon_2)^2 +4J^2}$ is the excitonic splitting, $r_{e_{\pm}}$ are the pumping rates of the energy eigenstates $|e_{\pm}\rangle$ and $ \gamma_{e_{\pm}}$ are incoherent emission rates. Table \ref{tab:table1} summarizes the relevant parameters introduced, and the density matrix elements $\{\rho_{e_+e_+},\rho_{e_-e_-},\rho^{R}_{e_+e_-},\rho^{I}_{e_+e_-}\}$ describe the populations of the DA energy-eigenstates and the real and imaginary parts of the coherences between these eigenstates, respectively.

We seek working master equations in the eigenstate basis, where the light-matter interaction is most easily described. To do this the matrices that appear in the Lindblad operators need to be rotated into the eigenbasis via the unitary matrix\cite{domcke_conical_2004,nitzan2013chemical}
\begin{equation}
\label{eq:rotata}
R(\theta) = \begin{pmatrix}
\cos(\theta) & -\sin(\theta)  \\
\sin(\theta) & \cos(\theta)
\end{pmatrix},
\end{equation}
where $\theta$ is the diabatic mixing angle given by
\begin{equation}
\tan(2\theta) = \frac{2J}{\epsilon_1 - \epsilon_2},
\end{equation}
where $0 \le \theta \le \frac{\pi}{2}$.

Four cases are relevant: local (in the site basis of the DA dimer) dephasing with either localized or delocalized trapping and global (in the energy-eigenbasis of the dimer) dephasing with either type of trapping. Specifically, diverse results can be obtained from each set of conditions, which are distinct from the often-studied localized trapping\cite{Wu2012,Caopreprint} and localized dephasing case. The two trapping cases considered correspond to two limiting regimes. The local trap arises when the dimer and the trap are close together allowing for spatial coherences to form between them. The delocalized trap corresponds to the opposite situation, in which there is a large spatial separation of the dimer and the trap that limits the formation of these coherences.

Consider first when dephasing is applied locally\cite{LenMontiel2014} (in the site basis of the DA dimer). Combining the results of Eqs (\ref{eq:Ldephase})-(\ref{eq:rotata}) with Eq.\ (\ref{ed:MEE}) the density matrix in this scenario evolves according to the following master equations:
\begin{eqnarray}
\label{eq:siteMEs}
\dot{\rho}_{e_{+}e_{+}} & = & -[2\Gamma + \gamma_d\sin^2(2\theta)+2\Gamma_{RC}\cos^2(\theta)]\rho_{e_{+}e_{+}} +\gamma_d\sin^2(2\theta)\rho_{e_{-}e_{-}} + r_{e_{+}}\rho_{gg} \nonumber  \\ 
& & +[\kappa\Gamma_{RC}\sin(2\theta) - 2\gamma_d\sin(2\theta)\cos(2\theta)]\rho^{R}_{e_+e_-}, \nonumber \\
\dot{\rho}_{e_{-}e_{-}} & = & -[2\Gamma + \gamma_d\sin^2(2\theta)+2\Gamma_{RC}\sin^2(\theta)]\rho_{e_{-}e_{-}} +\gamma_d\sin^2(2\theta)\rho_{e_{+}e_{+}} + r_{e_{-}}\rho_{gg} \nonumber  \\ 
& & +[\kappa\Gamma_{RC}\sin(2\theta) + 2\gamma_d\sin(2\theta)\cos(2\theta)]\rho^{R}_{e_+e_-},  \\
\dot{\rho}^{R}_{e_{+}e_{-}} & = & -[2\Gamma + \Gamma_{RC} + 2\gamma_d(1-\sin^2(2\theta))]\rho^{R}_{e_{+}e_{-}} +\Delta{\rho}^{I}_{e_{+}e_{-}} + \sqrt{r_{e_{+}} r_{e_{-}}}\rho_{gg} \nonumber \\
& & +\frac{\kappa}{2}\Gamma_{RC}\sin(2\theta)(\rho_{e_{+}e_{+}}+ \rho_{e_{-}e_{-}}) + \gamma_d\sin(2\theta)\cos(2\theta)(\rho_{e_-{e_-}}-\rho_{e_+{e_+}}),\nonumber \\
\dot{\rho}^{I}_{e_{+}e_{-}} & = & -[2\Gamma + 2\gamma_d + \Gamma_{RC}]\rho^{I}_{e_{+}e_{-}}  -\Delta\rho^{R}_{e_{+}e_{-}} ,\nonumber 
\end{eqnarray}
where the ground state population $\rho_{gg}$ remains close to unity and where any eigenstate population density matrix elements multiplied by the pumping rates $r_{e_{\pm}}$ or the spontaneous emission rates $\gamma_{e_{\pm}}$ are negligible. Both of these approximations are justified due to the extremely weak interactions of the incoherent light with the system. $\kappa$ is a parameter that is set to unity to describe localized trapping or to zero for delocalized trapping. Physically, this can be viewed as an alignment factor between the eigenstates and the reaction center. [Note the extension from Ref.\ \citenum{Tscherbul2018} where the DA system in that case was symmetric giving $r_{e_+}=r_{e_-}=r$. Here the rotation from the site basis to the eigenbasis modifies the pumping rates to $r_{e_+} = 2r\sin^2\theta$ and $r_{e_-} = 2r\cos^2\theta $.]

\begin{table}[h!]
  \begin{center}
    \caption{Parameter definitions}
    \label{tab:table1}
    \begin{tabular}{lc} % <-- Alignments: 1st column left, 2nd middle and 3rd right, with vertical lines in between
      \hline
      \hline
     $\Gamma$ & exciton recombination rate  \\
      $\gamma_d$ & phonon bath dephasing rate  \\
      $\Delta$  & excitonic splitting \\
      $r_i$  & incoherent light absorption rates \\
      $\Gamma_{RC}$  & \: reaction center excitonic trapping rate \\
      $\kappa$  & trapping mechanism switch \\
      \hline
       \hline
    \end{tabular}
  \end{center}
\end{table}

The second major case considered is when dephasing is applied globally (in the eigenbasis of the dimer). The system evolves similarly to Eq. (\ref{eq:siteMEs}) except now the dephasing rate affects both of the coherences equally and is absent in the evolution of the populations.\cite{volkhardmay2011} The master equations to describe this situation are given by
\begin{eqnarray}
\label{eq:eigenMEs}
\dot{\rho}_{e_{+}e_{+}} & = & -[2\Gamma +2\Gamma_{RC}\cos^2(\theta)]\rho_{e_{+}e_{+}} + r_{e_{+}}\rho_{gg} +\kappa\Gamma_{RC}\sin(2\theta)\rho^{R}_{e_+e_-}, \nonumber \\
\dot{\rho}_{e_{-}e_{-}} & = & -[2\Gamma + 2\Gamma_{RC}\sin^2(\theta)]\rho_{e_{-}e_{-}} + r_{e_{-}}\rho_{gg} +\kappa\Gamma_{RC}\sin(2\theta)\rho^{R}_{e_+e_-}, \nonumber \\
\dot{\rho}^{R}_{e_{+}e_{-}} & = & -[2\Gamma + \Gamma_{RC} + 2\gamma_d]\rho^{R}_{e_{+}e_{-}} +\Delta{\rho}^{I}_{e_{+}e_{-}} + \sqrt{r_{e_{+}} r_{e_{-}}}\rho_{gg} \\
& & +\frac{\kappa}{2}\Gamma_{RC}\sin(2\theta)(\rho_{e_{+}e_{+}}+ \rho_{e_{-}e_{-}}) + \gamma_d\sin(2\theta)\cos(2\theta)(\rho_{e_-{e_-}}-\rho_{e_+{e_+}}), \nonumber \\
\dot{\rho}^{I}_{e_{+}e_{-}} & = & -[2\Gamma + 2\gamma_d + \Gamma_{RC}]\rho^{I}_{e_{+}e_{-}}  -\Delta\rho^{R}_{e_{+}e_{-}}.\nonumber 
\end{eqnarray}
The relationships between the density matrix elements in the energy-eigenbasis and in the site basis are given by
\begin{equation}
\label{eq:rotate}
\rho_{11} = \sin^2\theta\rho_{e_+e_+} + \cos^2\theta\rho_{e_-e_-} + \sin(2\theta) \rho^R_{e_+e_-}, \nonumber
\end{equation}
\begin{equation}
\rho_{22} = \cos^2\theta\rho_{e_+e_+} + \sin^2\theta\rho_{e_-e_-} - \sin(2\theta) \rho^R_{e_+e_-} ,
\end{equation}
\begin{equation}
\rho_{12} = \frac{1}{2}\sin(2\theta)(\rho_{e_+e_+} - \rho_{e_-e_-}) + \cos(2\theta) \rho^R_{e_+e_-} -i\rho^I_{e_+e_-}. \nonumber
\end{equation}
For the symmetric case $\epsilon_1 = \epsilon_2$ the transformation reduces to:
\begin{equation}
\rho_{11} = \frac{1}{2}(\rho_{e_+e_+} + \rho_{e_-e_-} )+  \rho^R_{e_+e_-} , \nonumber
\end{equation}
\begin{equation}
\rho_{22} = \frac{1}{2}(\rho_{e_+e_+} + \rho_{e_-e_-} )-  \rho^R_{e_+e_-},
\label{eq:symcase}
\end{equation}
\begin{equation}
\rho_{12} = \frac{1}{2}(\rho_{e_+e_+} - \rho_{e_-e_-})  -i\rho^I_{e_+e_-}, \nonumber
\end{equation}
and the master equations given in Eqs. (\ref{eq:siteMEs}) and (\ref{eq:eigenMEs}) reduce to those of Eqs. (18) and (19) of Ref.\ \citenum{Tscherbul2018} respectively. [Note that there is an error in the population terms in Eq. (20) in Ref.\ \citenum{Tscherbul2018} which has been corrected here in Eq. (\ref{eq:symcase}).] It is interesting to note that the imaginary part of the coherence in the dimer system only changes in sign between the site and eigenstate basis and is independent of the rotation angle. We emphasize that this result is particular to dimer models and would not hold for more complex systems. We also take note that the master equations presented above only depend on $\epsilon_1$ and $\epsilon_2$ through $\Delta$ which contains the square of the energy difference. This has the effect that only $|\epsilon_1 - \epsilon_2|$ needs to be considered. i.e. the system is insensitive to the sign of the energy difference.

\subsection{Measure of energy transfer and exciton flux}
\label{sec:Measure}

The energy transfer efficiency is defined by the amount of population in the site neighboring the reaction center\cite{Manzano2013} appropriately weighted by the transfer rate to the reaction center and by the initial excitation rate. For localized trapping conditions it is given by
\begin{equation}
\eta_{\textrm{loc}} = \frac{\Gamma_{RC}}{r}\rho_{22} = \frac{\Gamma_{RC}}{r}\left(\cos^2\theta\rho_{e_+e_+} + \sin^2\theta\rho_{e_-e_-} - 2\sin\theta\cos\theta\rho^R_{e_+e_-}\right).
\label{eq:loceff}
\end{equation}
Equation (\ref{eq:loceff}) has a form suggestive of a coherent control scenario\cite{Shapiro2011} for unitary dynamics, i.e. two direct terms dependent on populations, and an interference-like term that is dependent on coherences. There is, however, a significant difference, emphasized later below. That is, in the unitary case the magnitude of the population terms do not depend on the coherences. This is not the case here, where the time dependence of the population terms is tied to the coherences via Eqs. (\ref{eq:siteMEs}) and (\ref{eq:eigenMEs}).
With this in mind $\eta_{\textrm{loc}}$ may be rewritten as
\begin{equation}
\eta_{\textrm{loc}} = \eta_{\textrm{direct}} + \eta_{\textrm{inter}},
\end{equation}
where $\eta_{\textrm{direct}}=\frac{\Gamma_{RC}}{r}(\cos^2\theta\rho_{e_+e_+} + \sin^2\theta\rho_{e_-e_-})$ and $\eta_{\textrm{inter}}=- 2\frac{\Gamma_{RC}}{r}\sin\theta\cos\theta\rho^R_{e_+e_-}$.
For the case of delocalized conditions the efficiency is given by
\begin{equation}
\eta_{\textrm{deloc}} = \frac{\Gamma_{RC}}{r}\left(\cos^2\theta\rho_{e_+e_+} + \sin^2\theta\rho_{e_-e_-}\right) =\eta_{\textrm{direct}},
\label{eq:deloceff}
\end{equation}
which reflects the fact that the coherences between different eigenstates have disappeared in this limiting case because of decoherence induced by delocalized trapping. Here, transfer happens directly between system eigenstates and the reaction center. This corresponds to an incoherent or Forster mechanism.\cite{Strmpfer2012,Sumi1999,Mukai1999,Scholes2001,Jang2004}

It is enlightening to explore the role of coherences in $\eta_{\textrm{loc}}$ in greater detail. For the pure dephasing model studied here the flux $F_{12}$ between the two sites is related to the imaginary component of the coherences (see Appendix A) as
\begin{equation}
\label{eq:Flux}
F_{12} = 2J\rho^{I}_{12} = -2J\rho^{I}_{e_+e_-},
\end{equation}
where the second equality follows from the third line of Eq.\ (\ref{eq:rotate}), which only holds for the two level systems. 

Equation (\ref{eq:Flux}) indicates that in an open quantum system subject to dissipation arising from pure dephasing conditions the flux between adjacent sites is entirely due to coherences. Interpreted in terms of $\rho_{12}$, this is a manifestation of spatial coherence between the two sites. 
That is, zero overlap of the spatial contributions of the two sites would yield $\rho_{12}=0$. A non-zero value indicates overlap. In addition, in the case of two sites, this is directly related to the ability of the energy eigenstates $|e_{\pm}\rangle$ to span both sites, reflected in $\rho^I_{e_+e_-}$. Further, $\rho^I_{e_+e_-}$ can be viewed as arising from interfering pathways, that is, it arises because $|e_+\rangle$ can emit into the background bath and $|e_-\rangle$ can absorb from the bath, creating two pathways to the bath. (Similarly, $|e_-\rangle$ can emit and $|e_+\rangle$ can absorb).

As shown in Appendix B Eqs. (\ref{eq:loceff}) and (\ref{eq:Flux}) can be related in the steady state limit through the use of the general relationship
\begin{equation}
\label{eq:ratio}
\frac{\rho^{R}_{e_+e_-}}{\rho^{I}_{e_+e_-}} = -\frac{2\Gamma + \Gamma_{RC}+2\gamma_d}{\Delta},
\end{equation}
which is valid regardless of the trapping conditions, the basis the dephasing is applied in, the pumping rate, and the asymmetry of the dimer. Notice that Eq.\ (\ref{eq:ratio}) implies that one of the coherence terms in the energy-eigenbasis is linearly dependent on the other. Using Eq.\ (\ref{eq:ratio}) with Eqs.\ (\ref{eq:loceff}) and (\ref{eq:Flux}) the localized trapping efficiency can be rewritten as
\begin{equation}
\eta_{\textrm{loc}} =\frac{\Gamma_{RC}}{r}\left(\cos^2\theta\rho_{e_+e_+} + \sin^2\theta\rho_{e_-e_-} - \sin(2\theta) \frac{2\Gamma + \Gamma_{RC}+2\gamma_d}{2J\Delta}F_{12}\right),
\label{eq:loceff_flux}
\end{equation}
which shows that the trapping efficiency is directly related to the site-to-site flux.

The character of Eq. (\ref{eq:loceff_flux}) is now clear. The first two terms are direct terms and the last term is an interference term. The magnitude of the interference contribution is modulated by the sum of the bath-induced dissipation rates, namely $(2\Gamma + \Gamma_{RC}+2\gamma_d)$, and system characteristics $(2J\Delta)$. Additional system dependence enters through the $\sin(2\theta)$ term and the $J$ contained within $F_{12}$. The flux  contains the explicit coherence contribution. 
A further comment should be made about the form of Eq. (\ref{eq:loceff_flux}). While it is tempting to refer to the first two terms as a classical contribution since these terms involve the populations, and the third term as a quantum contribution since the flux is related to the coherences, this attribution needs to be qualified since the steady state limit reached depends on the coupling of the population to the coherences, i.e., how Eqs. (\ref{eq:siteMEs}) and (\ref{eq:eigenMEs}) are constructed. Thus, there is always an implicit dependence on the coherences in the first two terms in Eq. (\ref{eq:loceff_flux}) and in Eq. (\ref{eq:deloceff}). This implicit dependence will be seen shortly for a series of limiting cases and explored further in the next section.

While it is possible to obtain analytic steady state solutions for Eqs. (\ref{eq:siteMEs}) and (\ref{eq:eigenMEs}), the general forms are too cumbersome to extract physically meaningful results so we focus on special cases to obtain insight into the behavior of the efficiency expressions. The regime of easily interpretable results is that in which $\Gamma_{RC} \gg \gamma_d$ (for details see Appendix C). That is,, for $\Gamma_{RC} \gg \gamma_d$:

(1) The first case of interest is that of a symmetric dimer with local dephasing and local trapping. In this limit the efficiency can be shown to be
\begin{eqnarray}
\label{eq:loceff_localdephase}
\eta_{\textrm{loc}} &=& \frac{\Gamma_{RC}}{r}\left(\rho_{e_+e_+} - \frac{\Gamma_{RC}}{\Delta^2}F_{12}\right), \nonumber \\
& = & \frac{\Gamma_{RC}}{r}\left(\frac{r}{\Gamma_{RC}}+ \frac{\Gamma_{RC}}{\Delta^2}F_{12} - \frac{\Gamma_{RC}}{\Delta^2}F_{12}\right), \\
& = & 1 \nonumber.
\end{eqnarray}
While the final result itself is uninteresting, as the dependence of $\eta_{\textrm{loc}}$ on any of the parameters has vanished, the second line in Eq. (\ref{eq:loceff_localdephase}) is enlightening as it shows that the flux contribution to the population term is exactly canceled out by the flux contribution from the interference term. Even though this is a special case it highlights the dependence of the steady state population values on the flux/coherences and shows that the one cannot just look at $\eta_{inter}$ to determine the role that the flux plays.

(2) For the case of a general dimer with local dephasing and delocalized trapping the efficiency is given by
\begin{equation}
\label{eq:deloceff_localdephase}
\eta_{\textrm{deloc}} =1,
\end{equation}
i.e., the efficiency is also unity and is also independent of $|\epsilon_1-\epsilon_2|$. In this limit the dependence of the efficiency on $\gamma_d$ cannot be inferred and Eq.\ (\ref{eq:deloceff_localdephase}) should be treated as an upper limit as it is an idealized case where the bath has minimal effects and the trapping rate is dominant. 

(3) Considering again the symmetric dimer, but now with global dephasing and local trapping, the efficiency can be expressed as
\begin{eqnarray}
\label{eq:loceff_globaldephase}
\eta_{\textrm{loc}} &=& \frac{F_{12}}{r} \left(\frac{1}{2} - \frac{\Gamma^2_{RC}}{\Delta^2}\right),\nonumber \\
&=& 1-\frac{2\Gamma^2_{RC}}{\Delta^2}
\end{eqnarray}
revealing that, in a similar fashion to Eq. (\ref{eq:loceff_localdephase}), the flux term (which is equal to $2r$) is canceled out and does not contribute in this limiting case. 

(4) The final case is that of a general dimer with global dephasing and delocalized trapping, and the efficiency is given by
\begin{equation}
\label{eq:deloceff_globaldephase}
\eta_{\textrm{deloc}} = \sin^2(2\theta),
\end{equation}
suggesting, along with Eq.\ (\ref{eq:deloceff_localdephase}), that in the Forster trapping regime the details of the dimer, other than the energy bias, doesn't have a large impact in the energy transfer efficiency when the trapping rate is faster than the bath dephasing rate.

A major limitation of the above results is that they do not reveal the dependence of the efficiency on the dephasing rate $\gamma_d$ or the flux. They are meant to serve as guides to understand some of the complex behavior within this model.
Hence, we now explore the stationary solutions of Eqs.\ (\ref{eq:siteMEs}) and (\ref{eq:eigenMEs}) numerically to study the dependence of Eqs. (\ref{eq:loceff_flux}) and (\ref{eq:deloceff}) outside of the parameter regimes where simple analytical results are obtainable.

\section{Numerical Results}
\label{sec:Results}

To obtain the stationary limit values of the density matrix either Eq.\ (\ref{eq:siteMEs})  or (\ref{eq:eigenMEs}) (depending on the context) was integrated using the RK4 method with initial conditions $\{\rho_{e_+e_+}(0),\rho_{e_-e_-}(0),\rho^{R}_{e_+e_-}(0),\rho^{I}_{e_+e_-}(0) \} = \{0,0,0,0 \}$ (i.e. $\rho_{gg}(0)=1$) with the remaining system parameters taken from Ref.\ \citenum{Tscherbul2018} to be: $\Gamma = 5\times10^{-4} \: \textrm{ps}^{-1}$, $\Gamma_{RC} = 0.5 \: \textrm{ps}^{-1}$, $J=0.12 \:\textrm{ps}^{-1}$, and $r = 6.34\times10^{-10} \: \textrm{ps}^{-1}$ until the density matrix values stop changing in time. The dephasing rate $\gamma_d$ and the DA energy difference are treated as adjustable parameters to study the competing effects of the phonon bath and energy gap on the energy transfer efficiency. Typical timescales for the steady state limit to be reached are within $10\textrm{ps}-100 \textrm{ps}$ depending on the dephasing rate.

\begin{figure}[t]
\centering
\includegraphics[width=0.8\textwidth]{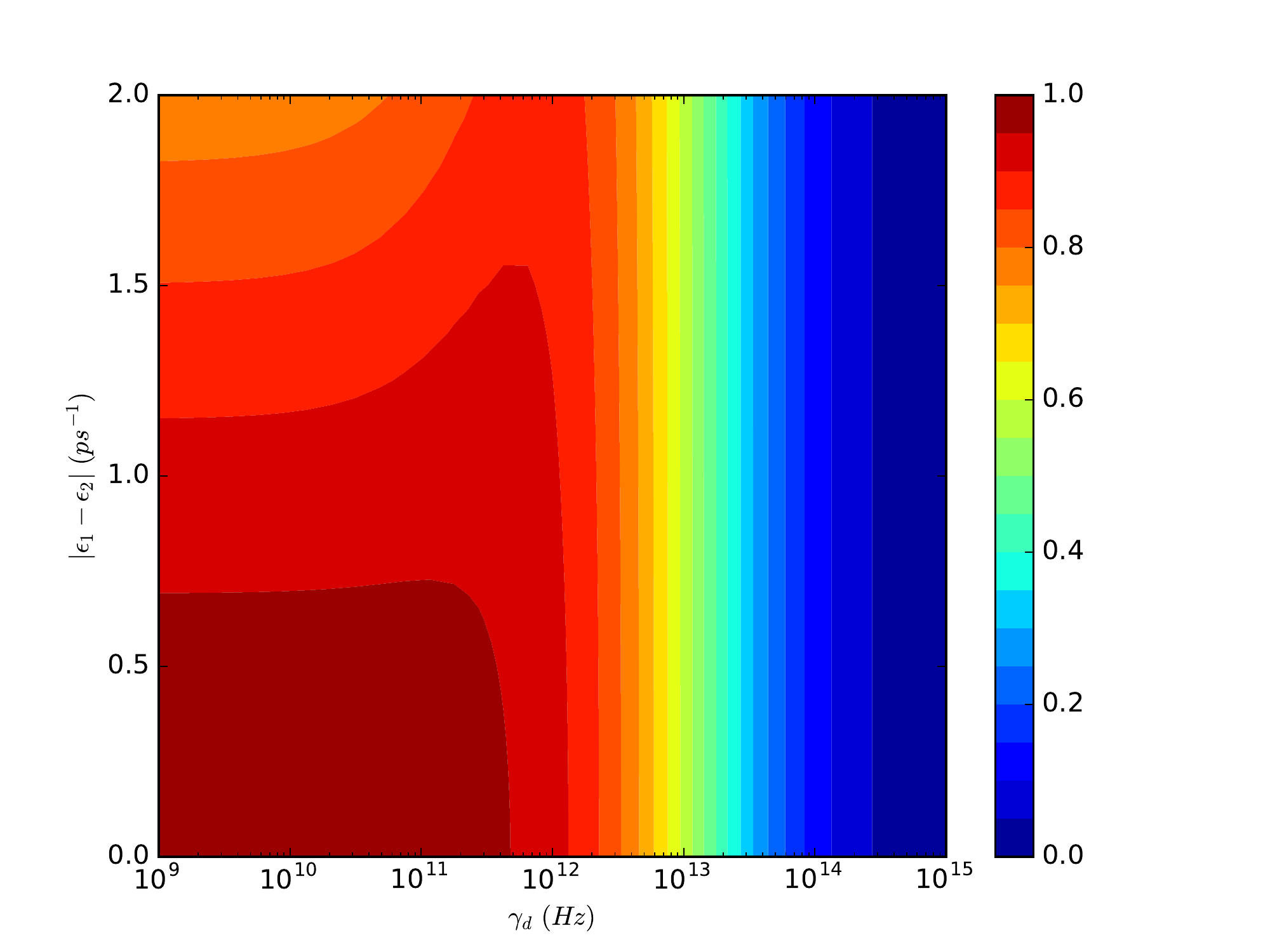} %figure name goes here
%\vspace*{-16mm}
\caption{Steady state values of $\eta_{\textrm{loc}}$ [Eq.\ (\ref{eq:loceff})] for localized trapping conditions as a function of the bath dephasing rate and the donor-acceptor energy splitting. Dephasing is applied locally.}
\protect\label{fig:1}
\end{figure}

\begin{figure}[t]
\centering
\includegraphics[width=0.8\textwidth]{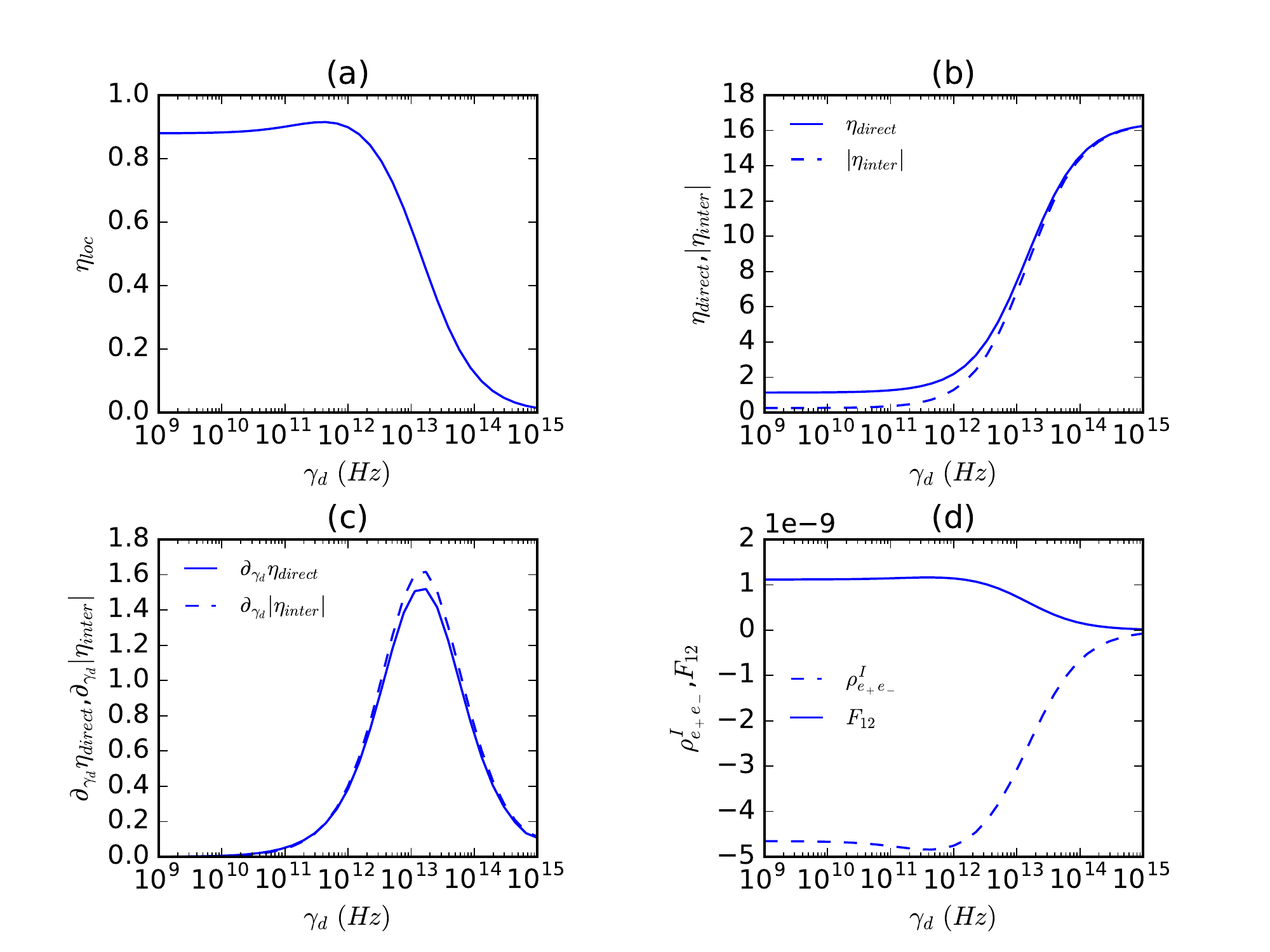} %figure name goes here
%\vspace*{-16mm}
\caption{A cut of Fig. \ref{fig:1} for $|\epsilon_1-\epsilon_2|=1.3 \: \textrm{ps}^{-1}$. (a) shows the efficiency as a function of $\gamma_d$. (b) the contributions from $\eta_{\textrm{direct}}$ and $\eta_{\textrm{inter}}$. (c) the derivative of each contribution. (d) the imaginary part of the coherence, and the flux. }
\protect\label{fig:2}
\end{figure}

Consider first the localized trapping condition case with dephasing applied locally. Figure \ref{fig:1} shows a contour plot of the steady state efficiency as a function of $\gamma_d$ and $|\epsilon_1-\epsilon_2|$. The value of $\epsilon_1-\epsilon_2=0$ corresponds to the symmetric dimer (i.e. $\Delta = 2J$) and as $|\epsilon_1-\epsilon_2|$ increases the two sites become more asymmetric. The efficiency shows a monotonic decrease as the dephasing rate increases for small $|\epsilon_1-\epsilon_2|$. When the absolute value of the DA energy difference reaches a value of $1.3 \: \textrm{ps}^{-1}$ the efficiency begins to show a turnover. This turnover in the efficiency is a unique behavior found only in asymmetric systems. An example of this turnover is presented in Fig. \ref{fig:2}, which also shows $\eta_{\textrm{direct}}$ and $\eta_{\textrm{inter}}$, to be discussed below. A similar turnover is seen in the imaginary part of the coherences. In fact, the enhancement of the efficiency appears when there is a minimum in $\rho^I_{e_+e_-}$,
corresponding to a maximum in the flux (see Eq.\ (\ref{eq:Flux})). Despite the difference in scale between Fig. 3(a) and Fig. 3(d) the flux and $\eta_{\textrm{loc}}$ appear to be proportional to each other. This maximum occurs due to an enchantment in the site-to-site flux arising from the environmental dephasing rate $\gamma_d$
\cite{Plenio2008,Chin2010,Wu2013,Rebentrost2009,Cao2009} and modulates the rate at which the interference grows. Prior to the maximum, $\eta_{\textrm{inter}}$ grows slower than $\eta_{\textrm{direct}}$ and is accompanied by a rising overall efficiency. After the maximum, $\eta_{\textrm{inter}}$ now grows faster approaching the value of $\eta_{\textrm{direct}}$ allowing the two terms to become competitive, leading to a vanishing efficiency. These effects are subtle, but demonstrative. If the dephasing is too slow the difference between the two contributions to the efficiency stays roughly constant due to the flux staying constant, and if the dephasing occurs too quickly the flux vanishes due to the coherence being destroyed between the two eigenstates which causes $\eta_{\textrm{inter}}$ to approach $-\eta_{\textrm{direct}}$.

\begin{figure}[h]
\centering
\includegraphics[width=0.8\textwidth]{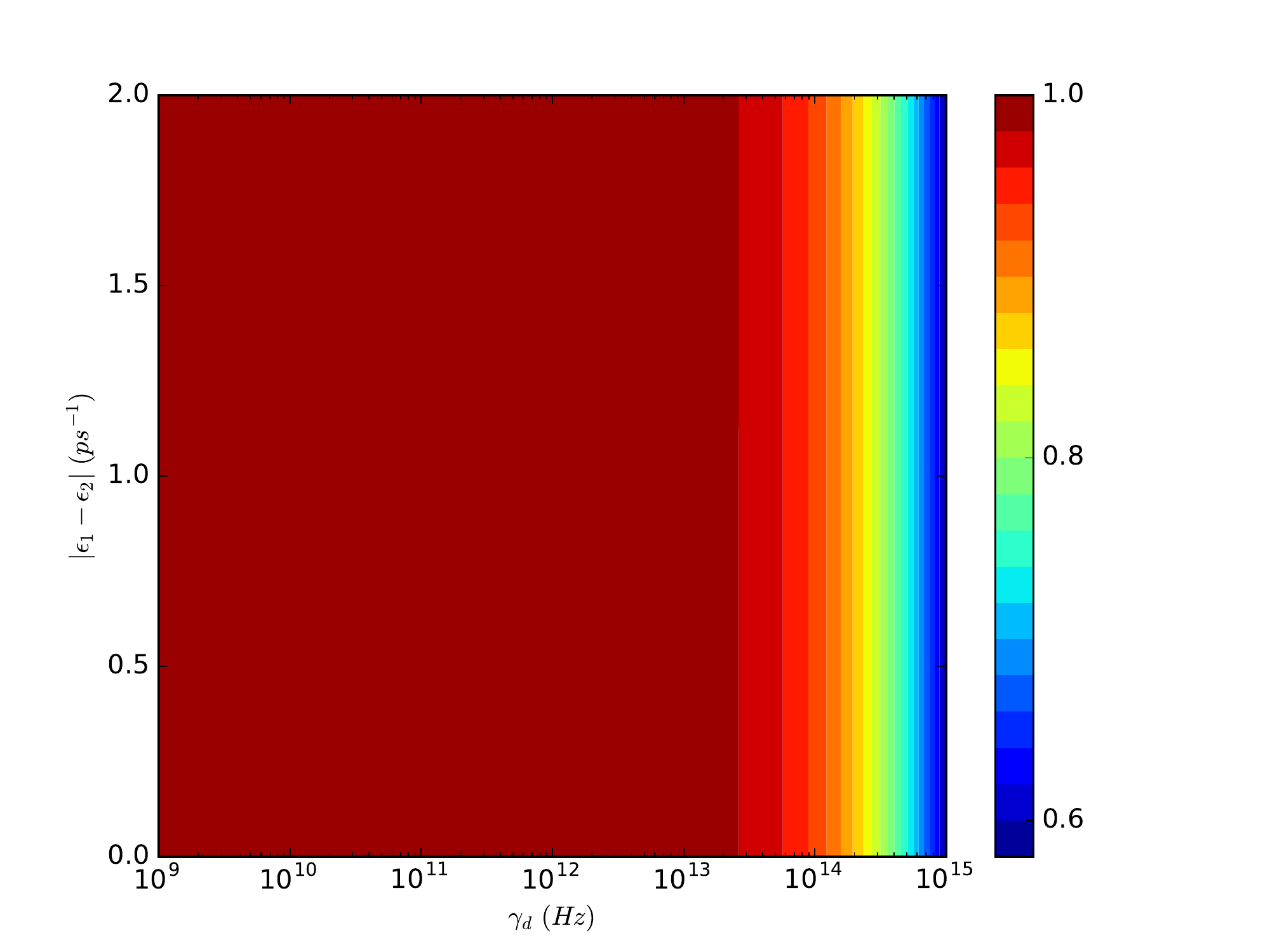} %figure name goes here
%\vspace*{-16mm}
\caption{Same as Fig.\ \ref{fig:1} but with $J=1.2$. Note the difference in scale between this figure and Fig. \ref{fig:1}.}
\protect\label{fig:largeJ}
\end{figure}

An important observation is that this turnover behavior is only found for small values of $J$ for the dephasing rates explored here. For example, Fig.\ \ref{fig:largeJ} shows that with increased $J$ the efficiency shows an overall increase in magnitude relative to Figure \ref{fig:1}. In this regime the system is so strongly coupled that dephasing has a much smaller effect on the flux as compared to the small $J$ case, and the rate at which the efficiency changes is monotonic, i.e. no maximum occurs.

%Figure 1
%--------------------------------------------------------------------------------------------
%--------------------------------------------------------------------------------------------
%Figure 2
%--------------------------------------------------------------------------------------------
%--------------------------------------------------------------------------------------------

%Figure largeJ
%--------------------------------------------------------------------------------------------
\begin{figure}[h]
\centering
\includegraphics[width=0.8\textwidth]{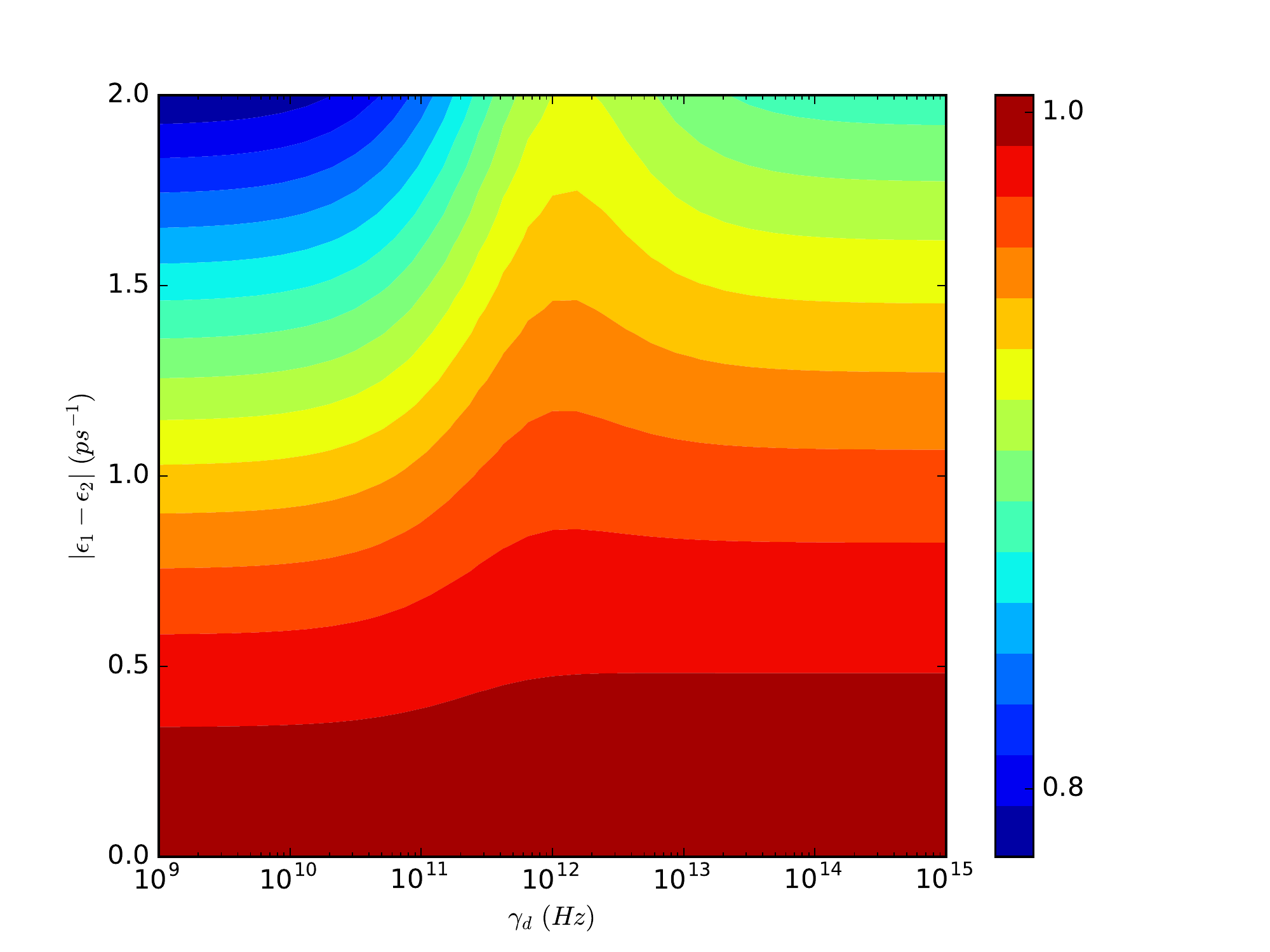} %figure name goes here
%\vspace*{-16mm}
\caption{Steady state values of $\eta_{\textrm{deloc}}$ [Eq.\ (\ref{eq:deloceff})] for delocalized trapping conditions as a function of the bath dephasing rate and the donor-acceptor splitting. Dephasing is applied locally.}
\protect\label{fig:3}
\end{figure}

\begin{figure}[h]
\centering
\includegraphics[width=0.8\textwidth]{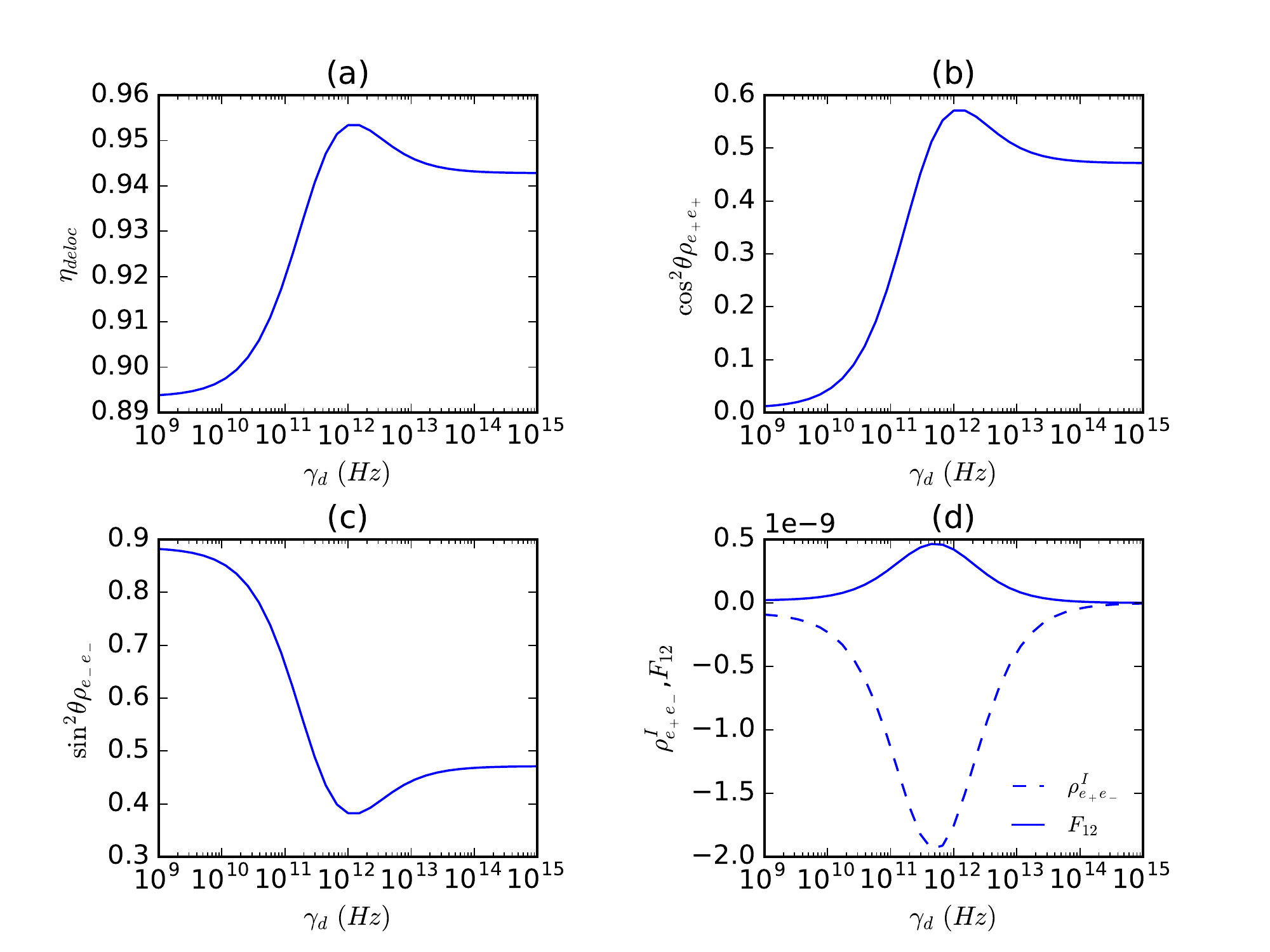} %figure name goes here
%\vspace*{-16mm}
\caption{A cut of Fig. \ref{fig:3} for $|\epsilon_1-\epsilon_2|=1.3  \: \textrm{ps}^{-1}$. (a) shows the efficiency as a function of $\gamma_d$. (b) and (c) the two contributions from $\eta_{\textrm{direct}}$. (d) the imaginary part of the coherences, and the flux.}
\protect\label{fig:35}
\end{figure}

%--------------------------------------------------------------------------------------------
Focusing now on the dimer with a delocalized trapping mechanism and dephasing applied locally we see from Fig. \ref{fig:3} a notable difference from the previous case in that the long term efficiency  for small splittings under this mechanism is almost completely insensitive to the magnitude of the dephasing rate and only depends on the splitting of the two eigenstates. This agrees with the result in Eq.\ (\ref{eq:deloceff_localdephase}) which states that the efficiency should be at or near unity.  Again, around $|\epsilon_1-\epsilon_2 |= 1.3 \: \textrm{ps}^{-1}$ the same enhancement in the efficiency is observed for similar values of $\gamma_d$. Figure \ref{fig:35} shows the same relation of the imaginary part of the coherences and the transfer efficiency as was seen in Fig. \ref{fig:2}. Here, the turnover is manifest in an interplay of the two contributions in $\eta_{\textrm{direct}}$, for smaller values of $\gamma_d$ the $\cos^2\theta\rho_{e_+e_+}$ contribution grows at faster rate than the $\sin^2\theta\rho_{e_-e_-}$ contribution. The general relationship of the efficiency and the flux/coherences is not as clearcut for this case as it was for the previous one. That is, as mentioned previously, although Eq.\ (\ref{eq:deloceff}) doesn't explicitly depend on the flux/coherences their presence is still seen in the modulation of the efficiency through the individual direct contributions. This implicit dependence is stronger in this case than the previous one and could be due to the fact that the $\kappa\Gamma_{RC}\sin(2\theta)$ terms are absent in Eq.\ (\ref{eq:siteMEs}) when $\kappa=0$, which could allow for more coherent behavior to manifest in $\eta_{\textrm{direct}}$ due to the absence of this dissipative term.
%Figure 3
%--------------------------------------------------------------------------------------------
%--------------------------------------------------------------------------------------------

%Figure 35
%--------------------------------------------------------------------------------------------
%--------------------------------------------------------------------------------------------

%Figure 4
%--------------------------------------------------------------------------------------------
\begin{figure}[h]
\centering
\includegraphics[width=0.8\textwidth]{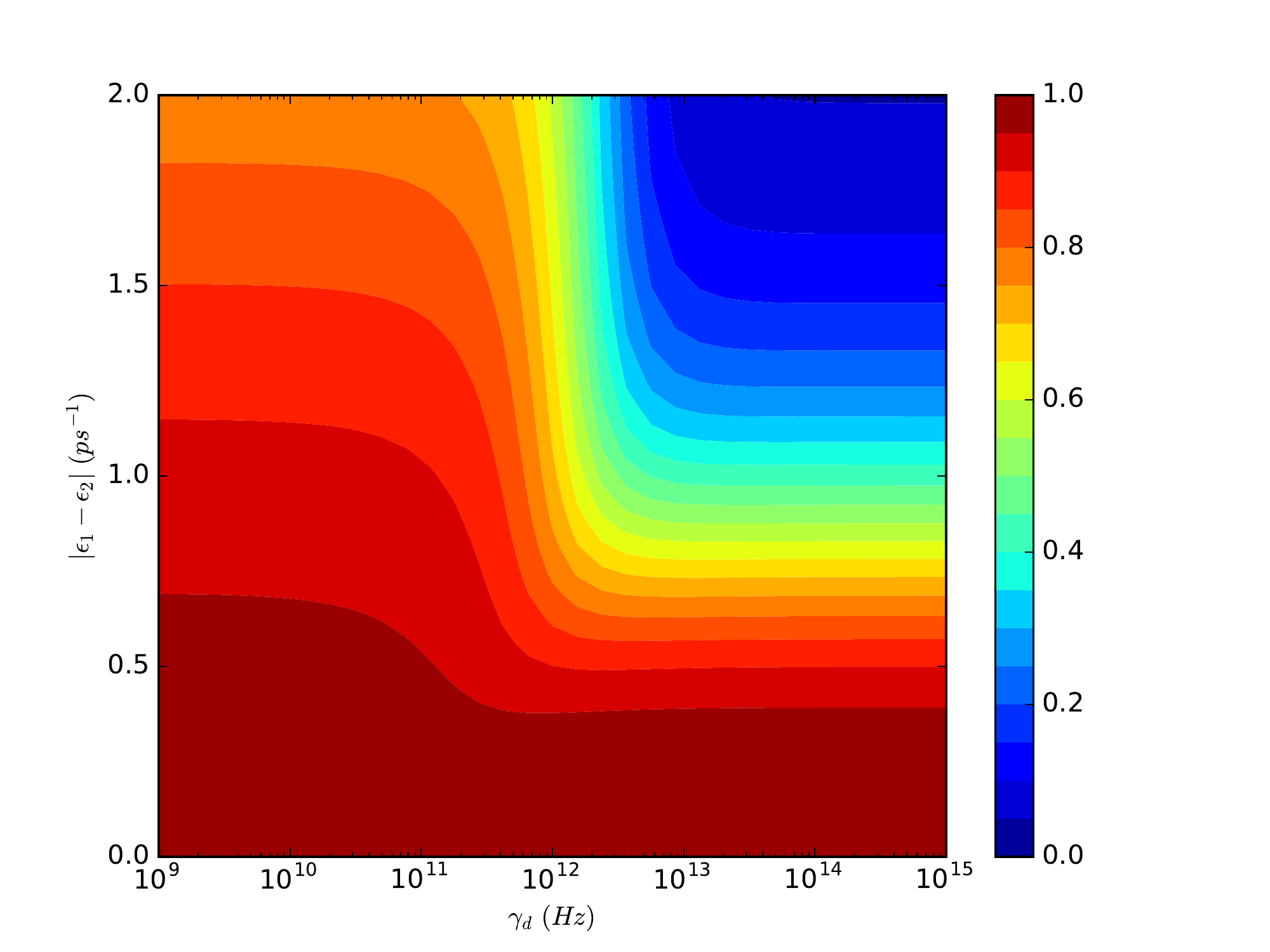} %figure name goes here
%\vspace*{-16mm}
\caption{Steady state values of $\eta_{\textrm{loc}}$ [Eq.\ (\ref{eq:loceff})] for localized trapping conditions as a function of the bath dephasing rate and the donor-acceptor splitting. Dephasing is applied globally.}
\protect\label{fig:4}
\end{figure}
%--------------------------------------------------------------------------------------------

Turning attention to localized trapping conditions when dephasing is applied globally, Fig. \ref{fig:4} demonstrates that the turnover behavior observed in the previous cases discussed is no longer present. In this set of conditions the efficiency monotonically decreases with increasing $\gamma_d$. For larger values of $|\epsilon_1-\epsilon_2|$ the efficiency behaves opposite to the case presented Fig. \ref{fig:3}, i.e. as the dephasing rate increases so does the efficiency for the delocalized trapping and local dephasing case, where here the reverse trend is seen. This effect would not be revealed if one focused on symmetric or nearly symmetric dimers as the efficiency is nearly independent of $\gamma_d$ for small site energy differences in both cases. This suggests that the trapping mechanism can play a significant role in modulation of the efficiency as the asymmetry of the sites increases.

%Figure 5
%--------------------------------------------------------------------------------------------
\begin{figure}[t]
\centering
\includegraphics[width=0.8\textwidth]{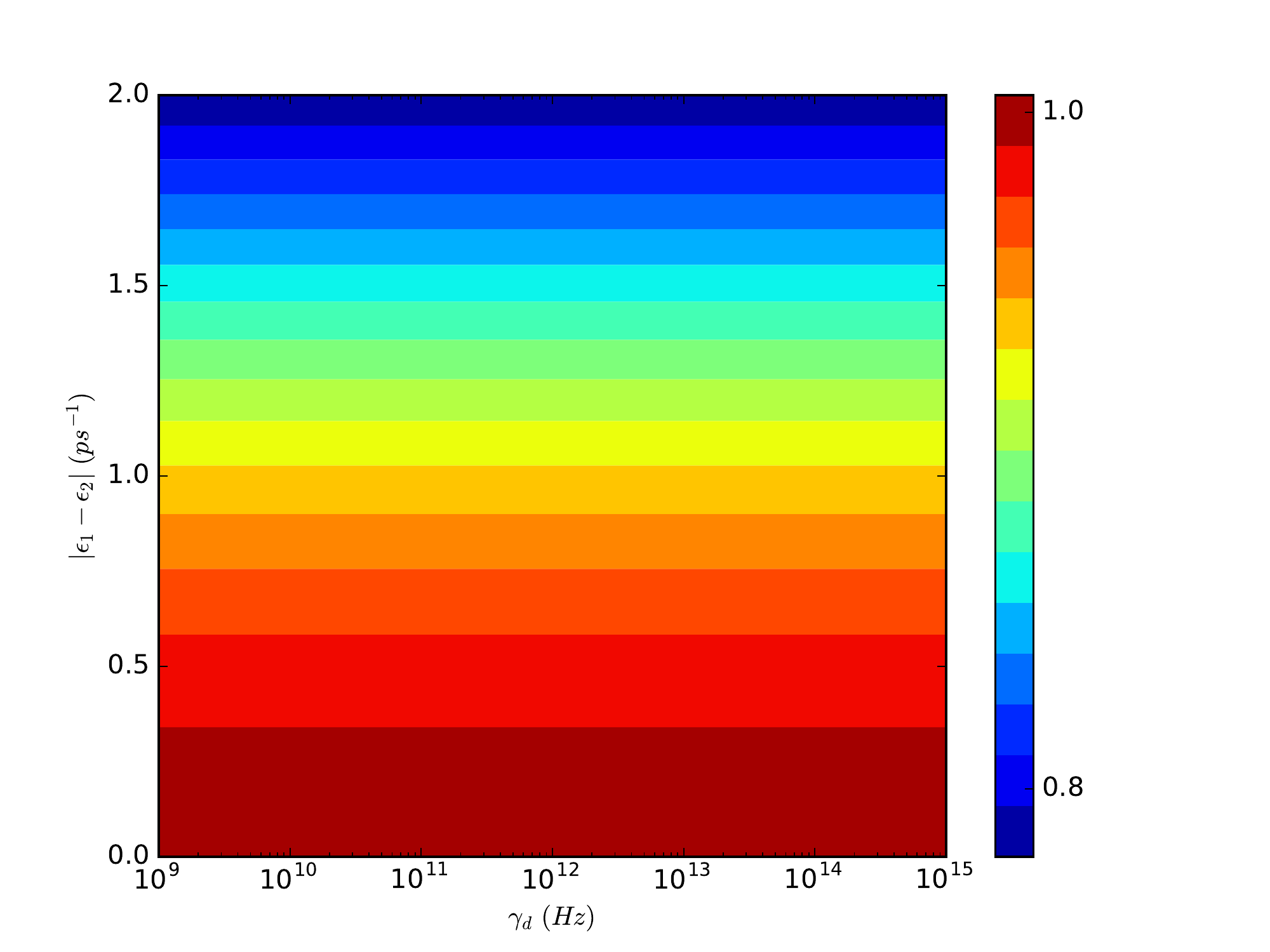} %figure name goes here
%\vspace*{-16mm}
\caption{Steady state values of $\eta_{\textrm{deloc}}$ [Eq.\ (\ref{eq:deloceff})] for delocalized trapping conditions as a function of the bath dephasing rate and the donor-acceptor splitting. Dephasing is applied globally.}
\protect\label{fig:5}
\end{figure}
%--------------------------------------------------------------------------------------------

The final case of global dephasing applied to a dimer with a delocalized trapping mechanism is shown in Fig. \ref{fig:5}. Similar to the case of the dimer with a delocalized trapping mechanism and local dephasing (Fig. \ref{fig:3}) the efficiency depends on the DA energy difference more than any other parameter. However, a unique feature of this case is that it appears that the efficiency \textit{only} depends on the site energy difference. This behavior is in excellent agreement within a factor of proportionality to Eq.\ (\ref{eq:deloceff_globaldephase}), which is rather surprising given that Eq.\ (\ref{eq:deloceff_globaldephase}) was derived for a special case. Here, then, the efficiency for this case is completely determined by the site energy difference.

\section{Conclusions}
\label{sec:Con}

We have generalized a set of master equations describing a dimer coupled to an environment comprised of a phonon bath and natural incoherent light to consider the effects of asymmetry on the energy transfer efficiency. It was seen that asymmetry reduces the overall efficiency as compared to the fully symmetric case regardless of trapping mechanism or dephasing model. Further, the results only depend upon $|\epsilon_1-\epsilon_2|$ as opposed to $\epsilon_1-\epsilon_2$.
We have also derived a general relationship between the ratio of the real and imaginary part of the coherences in the steady state limit. This result has allowed us to directly connect a contribution of $\eta_{\textrm{loc}}$ to the site-to-site flux and to the coherences. For a series of limiting cases we were able to relate the flux to the steady state population values, although in these cases the flux dependence was canceled out. We then went on to numerically demonstrate the role flux and bath induced dephasing play in energy transfer process for more general parameter regimes. Our results demonstrate that within the localized trapping models the coherence is closely related to the steady state efficiency. This suggests that if one finds a way to control and tune the coherences that the efficiency can likewise be controlled.

Under both local dephasing models it was found that a turnover behavior appears in the efficiency as a function of the dephasing rate $\gamma_d$ as the difference in site energies of the dimer become larger, regardless of the assumed trapping mechanism. This effect can only occur in asymmetric systems and reveals that in some cases there is an optimal amount of dephasing required to reach maximum efficiency for a given splitting. The results suggest that a turnover occurs regardless of the distance between the dimer and the reaction center. Rather, the importance lies in whether the bath couples to the individual dimer sites or to the eigenstates of the dimer. Further, we found that whenever this turnover occurred it was accompanied by a minimum in the steady state value of the imaginary part of the coherences or equivalently as a maximum in the flux. This behavior even appeared in the case of local dephasing and delocalized trapping in which the efficiency does not explicitly depend on the flux/coherence (see Eq.\ (\ref{eq:deloceff})) revealing an implicit dependence of the coherences on the steady state values of the eigenstate populations. This finding highlights the fact that a careful analysis is necessary when considering a delocalized (Forster) trap to asses whether the coherences play a role in the steady state efficiency. A summary of results is presented in Table II.

One key issue that needs to be overcome in future modeling is the use of pure dephasing rates. The Lindblad operators from which these dephasing rates stem generally have little physical meaning as they have no origin from a proper Hamiltonian and they also introduce many phenomenological parameters. Future work in this area must include a model Hamiltonian from which specific information can be derived. To address these shortcomings more rigorous modeling would need to be done to develop a deeper understanding of energy transfer and the role of asymmetry and coherences. Such work is ongoing in our laboratory.

The results contained here demonstrate that within the localized trapping models the coherence is closely related to the steady state efficiency. As a consequence, the efficiency can be controlled by
tuning the coherences, This is a direction that warrants further investigation.

\begin{table}[h!]
\begin{center}
    \caption{Summary of observations for the asymmetric dimer}
    \label{tab:observatio_nsummary}
\begin{tabular}{ | p{3cm} || p{6cm}|| p{6cm} | } 
\hline
\diagbox[width=8.2em]{Trapping}{Dephasing} & \hspace{5em}Local (site) & \hspace{5em}Global (eigen)  \\ 
\hline
\hline
\vspace{3.5em}Local & -turnover in the efficiency is observed as a function of $\gamma_d$ \:\: \:\:\:\:\: \: \: \: \:\:\: \: -efficiency is proportional to the site-to-site flux
\:\: \:\:\: \:\:\: \:\:\: \:\:\:\:\: \:\:\: \: \: \:\:\: \: \: \:\:\: \: \:\:\: \: \: \:\:\: \: -strong dependence on $\gamma_d$ & -no turnover behavior is observed
\:\: \:\:\: \:\:\: \:\:\: \:\:\:\:\: \:\:\: \: \:\:\:  \: \:\:\: \:\:\: \:\: \: \: \:\:\: \: -strong dependence on $\gamma_d$ \\ 
\hline
\vspace{3.5em}Delocalized (Forster) & -turnover in the efficiency is observed as a function of $\gamma_d$ \:\: \:\:\:\:\: \: \: -efficiency has a weaker flux dependence than in the local trapping cases
\:\: \:\:\: \:\:\: \:\:\: \:\:\:\:\: \:\:\:\:\: \: \:\:\: \:\:\: \: \:\:\: \: \: \:\:\: \: -weaker dependence on $\gamma_d$ than in local trapping cases & -no turnover behavior is observed 
\:\: \:\:\: \:\:\: \:\:\: \:\:\:\:\: \:\:\: \: \:\:\: \: -the efficiency is independent of both the flux and $\gamma_d$ and only depends on $|\epsilon_1-\epsilon_2|$ \\ 
\hline
\end{tabular}
\end{center}
\end{table}

\section*{Acknowledgments}

We gratefully acknowledge support from the U.S. Air Force Scientific Research under grants FA9550-17-1-0310 and FA9550-19-1-0267.

\appendix

\section{Site-to-site flux under pure dephasing conditions}
\label{app:AFlux}

Issues of the relation of the flux to system coherences (Eq.\ (\ref{eq:Flux})) are often presented without a clear discussion of the conditions under which this is case. Here we derive this relationship, providing insight into the conditions under which Eq.\ (\ref{eq:Flux}) holds.

Consider the following exciton Hamiltonian
\begin{equation}
H = H_S + H_B + H_I,
\end{equation}
where
\begin{equation}
H_S = \sum_n \epsilon_n b^{\dagger}_n b_n + \sum_{mn} J_{mn} b^{\dagger}_n b_m,
\end{equation}
is the system Hamiltonian consisting of $n$ sites and $b_n =  | 0 \rangle\langle n |$ where $| 0 \rangle$ denotes the ground state and $| n \rangle$ a singly excited state. The bath and interaction Hamiltonians are given by
\begin{equation}
H_B = \sum_k \Omega_k a^{\dagger}_k a_k,
\end{equation}
and
\begin{equation}
H_I = \sum_{n} g_{n}b^{\dagger}_n b_n.
\end{equation}
Note that the interaction Hamiltonian is proportional to the occupation on the sites i.e. $H_I\propto | n \rangle\langle n |$. This is the pure dephasing condition assumed, for example, in the Lindblad operators.

The rate of change of the population in each site is given by
\begin{equation}
\langle\dot{b^{\dagger}_n b_n}\rangle =\langle \dot{P_n} \rangle = -i\langle  [H,b^{\dagger}_n b_n] \rangle = i\sum_{m\neq n}J_{mn}\langle \left(b^{\dagger}_m b_n -b^{\dagger}_n b_m \right) \rangle.
\label{eq:1}
\end{equation}
Since the total population is conserved i.e. $\sum_n P_n =1$ this implies the continuity condition
\begin{equation}
\langle \dot{P_n} \rangle = \sum_{m\neq n} F_{mn},
\label{eq:2}
\end{equation}
where $F_{mn}$ is the flux from site $m$ to site $n$. Comparing Eq.\ \ref{eq:1} and Eq.\ \ref{eq:2} we see that
\begin{equation}
F_{mn} = iJ_{mn}\langle \left(b^{\dagger}_m b_n -b^{\dagger}_n b_m \right) \rangle = 2J_{mn}\rho^{I}_{mn},
\label{eq:dephasFlux}
\end{equation}
which states that the flux from one site to another is solely dependent on the coherences between them. This statement is only true when $H_I\propto | n \rangle\langle n |$. To demonstrate this point consider changing the interaction Hamiltonian to 
\begin{equation}
H_I = \sum_{n} g_{n}(b^{\dagger}_n + b_n).
\end{equation}
Note that the interaction Hamiltonian is still dependent only on system operators but it no longer dependent on the population of each site. Repeating the calculation above with this new interaction form yields
\begin{equation}
F_{mn} = 2J_{mn} \rho^{I}_{mn} + ig_{n}\langle(b^{\dagger}_n - b_n)\rangle.
\end{equation}
Notice the additional term that has appeared which is a direct result of new form of $H_I$.

It is also worth noting that if the form of the interaction is taken to be
\begin{equation}
H_I = \sum_{kn} g_{nk}(a_k^{\dagger}+a_k)b^{\dagger}_n b_n,
\end{equation}
Then the flux will still be given by Eq.\ (\ref{eq:dephasFlux}) due to the fact that $Tr_B\left[\rho^{eq}(a_k^{\dagger}+a_k) \right]=0$ for a harmonic bath.

\section{Derivation of Eq.\ (\ref{eq:ratio})}
\label{app:Aratio}

We begin by rewriting Eq.\ (\ref{eq:siteMEs}) in matrix form as
\begin{equation}
\dot{\rho}(t) = \boldsymbol{A}\rho(t) +B,
\end{equation}
where $\rho(t) = [ \rho_{e_+e_+}(t), \rho_{e_-e_-}(t), \rho^{R}_{e_+e_-}(t) ,\rho^{I}_{e_+e_-}(t) ]$, $B=[r_{e_+},r_{e_-},\sqrt{r_{e_+}r_{e_-}},0]$, and
\begin{gather}
\boldsymbol{A}
 =
  \begin{bmatrix}
   -(\Gamma_+ + d) & d & \Gamma^*_{RC}  + 2\gamma^*_d & \:\: 0 \\
  d &  -(\Gamma_- + d) & \Gamma^*_{RC}  - 2\gamma^*_d & \:\: 0 \\
  \frac{1}{2}(\Gamma^*_{RC}  - 2\gamma^*_d) & \frac{1}{2}(\Gamma^*_{RC}  -+2\gamma^*_d) & -\gamma_s +2d & \:\: \Delta \\
  0 & 0 & -\Delta & \:\:-\gamma_s
   \end{bmatrix},
\end{gather}
with $\Gamma_+ = 2\Gamma  +2\Gamma_{RC}\cos^2\theta$, $\Gamma_- = 2\Gamma  +2\Gamma_{RC}\sin^2\theta$, $\Gamma^*_{RC} = k\Gamma_{RC}\sin(2\theta)$, $\gamma_s= 2\Gamma  +\Gamma_{RC} +2\gamma_d$, $\gamma^*_d=\gamma_d \sin(2\theta) \cos(2\theta)$, and $d= \gamma_d \sin^2(2\theta)$.
The steady state solutions can be obtained from
\begin{equation}
\rho(\infty) =- \boldsymbol{A}^{-1}B,
\end{equation}
which yields for the coherences
\begin{eqnarray}
\rho^{R}_{e_+e_-}(\infty) &=& -\frac{\gamma_s}{\textrm{det}(\boldsymbol{A})} \{r_{e_+}[-\Gamma_-\Gamma^*_{RC} + 2\Gamma_-\gamma^*_d -2\Gamma^*_{RC}d] \nonumber \\
& & + r_{e_-} [-\Gamma_+\Gamma^*_{RC} - 2\Gamma_+\gamma^*_d -2\Gamma^*_{RC}d] \nonumber \\
& & -2\sqrt{r_{e_+}r_{e_-}}[\Gamma_+\Gamma_- + \Gamma_+g + \Gamma_-d] \},
\end{eqnarray}
and
\begin{eqnarray}
\rho^{I}_{e_+e_-}(\infty) &=& \frac{\Delta}{\textrm{det}(\boldsymbol{A})} \{r_{e_+}[-\Gamma_-\Gamma^*_{RC} + 2\Gamma_-\gamma^*_d -2\Gamma^*_{RC}d] \nonumber \\
& & + r_{e_-} [-\Gamma_+\Gamma^*_{RC} - 2\Gamma_+\gamma^*_d -2\Gamma^*_{RC}d] \nonumber \\
& & -2\sqrt{r_{e_+}r_{e_-}}[\Gamma_+\Gamma_- + \Gamma_+g + \Gamma_-d] \},
\end{eqnarray}

\begin{equation}
\frac{\rho^{R}_{e_+e_-}(\infty)}{\rho^{I}_{e_+e_-}(\infty)} = -\frac{\gamma_s}{\Delta} = -\frac{2\Gamma + \Gamma_{RC}+2\gamma_d}{\Delta},
\end{equation}
which is Eq.\ (\ref{eq:ratio}). One can follow similar steps to verify that if Eq.\ (\ref{eq:eigenMEs}) were used instead of Eq.\ (\ref{eq:siteMEs}) then the end result would be the same, showing that the result is independent of how the dephasing is applied. Equation (\ref{eq:ratio}) is also independent of $\kappa$, meaning that the trapping conditions also do not change the ratio of the real and imaginary parts of the coherences in the steady state limit.

\section{Efficiency expressions in the limit of $\Gamma_{RC}\gg\gamma_d$}
\label{app:Effresults}

\textit{Equation (\ref{eq:loceff_localdephase})}: The steady state solutions of Eqs. (\ref{eq:siteMEs}) and (\ref{eq:eigenMEs}) in the limit of $\Gamma_{RC}\gg\gamma_d$ can be obtained in a similar manner as outlined in Appendix B. The results for Eq. (\ref{eq:siteMEs}) for the case of $\theta=\frac{\pi}{4}$ and $\kappa=1$ are given by
\begin{eqnarray}
\label{eq:Limitsols_siteloc}
\rho_{e_{\pm}e_{\pm}} & = & \frac{r}{\Gamma_{RC}} + \frac{2r\Gamma_{RC}}{\Delta^2} = \frac{r}{\Gamma_{RC}} + \frac{\Gamma_{RC}}{\Delta^2}F_{12},   \nonumber \\
\rho}^{R}_{e_{+}e_{-} & = & \frac{2r\Gamma_{RC}}{\Delta^2}, \\
\rho}^{I}_{e_{+}e_{-} & = & -\frac{2r}{\Delta}, \nonumber
\end{eqnarray}
where we have made use of the fact that $F_{12}=2r$ in the symmetric dimer to obtain the second equality of the first line. Substituting Eq. (\ref{eq:Limitsols_siteloc}) into Eq. (\ref{eq:loceff_flux}) yields Eq. (\ref{eq:loceff_localdephase}). 

\textit{Equation (\ref{eq:deloceff_localdephase})}: The solutions to Eq. (\ref{eq:siteMEs}) if one sets $\kappa=0$ are
\begin{eqnarray}
\label{eq:Limitsols_sitedeloc}
\rho_{e_{\pm}e_{\pm}} & = &  \frac{r_{e_{\pm}}}{\Gamma_{RC}}  \nonumber \\
\rho}^{R}_{e_{+}e_{-} & = & \Gamma_{RC}\frac{\sqrt{r_{e_+}r_{e_-}}}{\Gamma_{RC}^2+\Delta^2} \\
\rho}^{I}_{e_{+}e_{-} & = & -\Delta\frac{\sqrt{r_{e_+}r_{e_-}}}{\Gamma_{RC}^2+\Delta^2} \nonumber.
\end{eqnarray}
If one inserts Eq. (\ref{eq:Limitsols_sitedeloc}) into Eq. (\ref{eq:deloceff}) the outcome is Eq. (\ref{eq:deloceff_localdephase}). 

\textit{Equation (\ref{eq:loceff_globaldephase})}: Now turning to the solutions of Eq. (\ref{eq:eigenMEs}) when $\theta=\frac{\pi}{4}$ and $\kappa=1$, the results are
\begin{eqnarray}
\label{eq:Limitsols_eigenloc}
\rho_{e_{\pm}e_{\pm}} & = &  \frac{r}{\Gamma_{RC}} = \frac{F_{12}}{2\Gamma_{RC}}  \nonumber \\
\rho}^{R}_{e_{+}e_{-} & = & \frac{2r\Gamma_{RC}}{\Delta^2} \\
\rho}^{I}_{e_{+}e_{-} & = & -\frac{2r}{\Delta} \nonumber,
\end{eqnarray}
where, again, we have made use of the relationship $F_{12}=2r$. Substituting Eq. (\ref{eq:Limitsols_eigenloc}) into Eq. (\ref{eq:loceff_flux}) yields Eq. (\ref{eq:loceff_globaldephase}). 

\textit{Equation (\ref{eq:deloceff_globaldephase})}: In the case where $\kappa=0$ Eq. (\ref{eq:eigenMEs}) has the following solutions
\begin{eqnarray}
\label{eq:Limitsols_eigenloc}
\rho_{e_{\pm}e_{\pm}} & = &  \frac{r_{e_{\pm}}}{\Gamma_{\pm}}  \nonumber \\
\rho}^{R}_{e_{+}e_{-} & = & \frac{\Gamma_{RC}\sqrt{r_{e_+}r_{e_-}}}{\Gamma_{RC}^2+\Delta^2} \\
\rho}^{I}_{e_{+}e_{-} & = & -\Delta\frac{\sqrt{r_{e_+}r_{e_-}}}{\Gamma_{RC}^2+\Delta^2} \nonumber.
\end{eqnarray}
Equation (\ref{eq:deloceff_globaldephase}) is obtained if Eq. (\ref{eq:Limitsols_eigenloc}) is inserted into Eq. (\ref{eq:deloceff}).

%\bibliography{biblo}

\begin{thebibliography}{10}

\bibitem{Fleming2011}
G.~R. Fleming, G.~D. Scholes, and Y.-C. Cheng,
\newblock Procedia Chemistry {\bf 3}, 38 (2011).

\bibitem{Ball2011}
P.~Ball,
\newblock Nature {\bf 474}, 272 (2011).

\bibitem{Wu2012}
J.~Wu, F.~Liu, J.~Ma, R.~J. Silbey, and J.~Cao,
\newblock The Journal of Chemical Physics {\bf 137}, 174111 (2012).

\bibitem{Engel2007}
G.~S. Engel et~al.,
\newblock Nature {\bf 446}, 782 (2007).

\bibitem{Ishizaki2012}
A.~Ishizaki and G.~R. Fleming,
\newblock Annual Review of Condensed Matter Physics {\bf 3}, 333 (2012).

\bibitem{Jiang1991}
X.-P. Jiang and P.~Brumer,
\newblock The Journal of Chemical Physics {\bf 94}, 5833 (1991).

\bibitem{Brumer2018}
P.~Brumer,
\newblock The Journal of Physical Chemistry Letters {\bf 9}, 2946 (2018).

\bibitem{Manal2010}
T.~Man{\v{c}}al and L.~Valkunas,
\newblock New Journal of Physics {\bf 12}, 065044 (2010).

\bibitem{Tscherbul2018}
T.~V. Tscherbul and P.~Brumer,
\newblock The Journal of Chemical Physics {\bf 148}, 124114 (2018).

\bibitem{Manzano2013}
D.~Manzano,
\newblock {PLoS} {ONE} {\bf 8}, e57041 (2013).

\bibitem{Dodin2019}
A.~Dodin and P.~Brumer,
\newblock The Journal of Chemical Physics {\bf 150}, 184304 (2019).

\bibitem{Dodin2016}
A.~Dodin, T.~V. Tscherbul, and P.~Brumer,
\newblock The Journal of Chemical Physics {\bf 144}, 244108 (2016).

\bibitem{olina2014natural}
J.~Olina, A.~G. Dijkstra, C.~Wang, and J.~Cao,
\newblock Can natural sunlight induce coherent exciton dynamics?,
  ArXiv:1408.5385.

\bibitem{LenMontiel2014}
R.~de~J.~Le{\'{o}}n-Montiel, I.~Kassal, and J.~P. Torres,
\newblock The Journal of Physical Chemistry B {\bf 118}, 10588 (2014).

\bibitem{Pachn2017}
L.~A. Pach{\'{o}}n, J.~D. Botero, and P.~Brumer,
\newblock Journal of Physics B: Atomic, Molecular and Optical Physics {\bf 50},
  184003 (2017).

\bibitem{Jesenko2013}
S.~Jesenko and M.~{\v{Z}}nidari{\v{c}},
\newblock The Journal of Chemical Physics {\bf 138}, 174103 (2013).

\bibitem{Tscherbul2014}
T.~V. Tscherbul and P.~Brumer,
\newblock Physical Review Letters {\bf 113} (2014).

\bibitem{Tscherbul2015}
T.~V. Tscherbul and P.~Brumer,
\newblock The Journal of Chemical Physics {\bf 142}, 104107 (2015).

\bibitem{Kozlov2006}
V.~V. Kozlov, Y.~Rostovtsev, and M.~O. Scully,
\newblock Physical Review A {\bf 74} (2006).

\bibitem{Lindblad1976}
G.~Lindblad,
\newblock Communications in Mathematical Physics {\bf 48}, 119 (1976).

\bibitem{Breuer2007}
H.-P. Breuer and F.~Petruccione,
\newblock {\em The Theory of Open Quantum Systems},
\newblock Oxford University Press, 2007.

\bibitem{domcke_conical_2004}
W.~Domcke, D.~Yarkony, and H.~K{\"o}ppel, editors,
\newblock {\em Conical intersections: electronic structure, dynamics \&
  spectroscopy},
\newblock Number v.15 in Advanced series in physical chemistry, World
  Scientific, River Edge, NJ, 2004.

\bibitem{nitzan2013chemical}
A.~Nitzan,
\newblock {\em Chemical dynamics in condensed phases : relaxation, transfer,
  and reactions in condensed molecular systems},
\newblock Oxford University Press, Oxford, 2013.

\bibitem{Caopreprint}
P.-Y. Yang and J.~Cao,
\newblock preprint: Steady-state analysis of a photosynthetic dimer driven by
  incoherent light, Preprint,2020.

\bibitem{volkhardmay2011}
O.~K{\"u}hn and V.~May,
\newblock {\em Charge and Energy Transfer Dynamics in Molecular Systems},
\newblock Wiley-VCH, 2011.

\bibitem{Shapiro2011}
M.~Shapiro and P.~Brumer,
\newblock {\em Quantum Control of Molecular Processes},
\newblock Wiley-{VCH} Verlag {GmbH} {\&} Co. {KGaA}, 2011.

\bibitem{Strmpfer2012}
J.~Str\"{u}mpfer, M.~{\c{S}}ener, and K.~Schulten,
\newblock The Journal of Physical Chemistry Letters {\bf 3}, 536 (2012).

\bibitem{Sumi1999}
H.~Sumi,
\newblock The Journal of Physical Chemistry B {\bf 103}, 252 (1999).

\bibitem{Mukai1999}
K.~Mukai, S.~Abe, and H.~Sumi,
\newblock The Journal of Physical Chemistry B {\bf 103}, 6096 (1999).

\bibitem{Scholes2001}
G.~D. Scholes, X.~J. Jordanides, and G.~R. Fleming,
\newblock The Journal of Physical Chemistry B {\bf 105}, 1640 (2001).

\bibitem{Jang2004}
S.~Jang, M.~D. Newton, and R.~J. Silbey,
\newblock Physical Review Letters {\bf 92} (2004).

\bibitem{Plenio2008}
M.~B. Plenio and S.~F. Huelga,
\newblock New Journal of Physics {\bf 10}, 113019 (2008).

\bibitem{Chin2010}
A.~W. Chin, A.~Datta, F.~Caruso, S.~F. Huelga, and M.~B. Plenio,
\newblock New Journal of Physics {\bf 12}, 065002 (2010).

\bibitem{Wu2013}
J.~Wu, R.~J. Silbey, and J.~Cao,
\newblock Physical Review Letters {\bf 110} (2013).

\bibitem{Rebentrost2009}
P.~Rebentrost, M.~Mohseni, I.~Kassal, S.~Lloyd, and A.~Aspuru-Guzik,
\newblock New Journal of Physics {\bf 11}, 033003 (2009).

\bibitem{Cao2009}
J.~Cao and R.~J. Silbey,
\newblock The Journal of Physical Chemistry A {\bf 113}, 13825 (2009).

\end{thebibliography}
%\bibliographystyle{biblo}

\end{document}